\documentclass[twocolumn]{aastex631}
\usepackage{float}
\usepackage{graphicx}
\usepackage{amsmath}
\usepackage{natbib}
\usepackage{color}
\usepackage{verbatim}
\usepackage[utf8]{inputenc}
\usepackage{mathtools}
\usepackage{amssymb}
\usepackage{textcomp}
\usepackage{enumitem}
\usepackage{braket}
\usepackage{graphicx} 

\newcommand{\minus}{\scalebox{0.75}[1.0]{$-$}}

\begin{document}
\title{A Timeline of the M81 Group: Properties of the Extended Structures of M82 and NGC 3077}


\author[0009-0001-1147-6851]{Benjamin N.\ Velguth}
\affiliation{Department of Astronomy, University of 
Michigan, 323 West Hall, 1085 S. University Ave., Ann Arbor, MI, 48105-1107, USA} 

\author[0000-0002-5564-9873]{Eric F.\ Bell}
\affiliation{Department of Astronomy, University of 
Michigan, 323 West Hall, 1085 S. University Ave., Ann Arbor, MI, 48105-1107, USA} 

\author[0000-0003-2599-7524]{Adam Smercina}\thanks{Hubble Fellow}
\affiliation{Department of Astronomy, Box 351580, University of Washington, Seattle, WA 98195, USA}
\affiliation{Space Telescope Science Institute, 3700 San Martin Dr., Baltimore, MD 21218, USA}

\author[0000-0003-0511-0228]{Paul Price}
\affiliation{Department of Astrophysical Sciences, Princeton University, Princeton, NJ 08544, USA}

\author[0000-0003-2294-4187]{Katya Gozman}
\affiliation{Department of Astronomy, University of 
Michigan, 323 West Hall, 1085 S. University Ave., Ann Arbor, MI, 48105-1107, USA} 

\author[0000-0003-2325-9616]{Antonela Monachesi}
\affiliation{Departamento de Astronom\'ia, Universidad de La Serena, Av. Ra\'ul Bitr\'an 1305, La Serena, Chile}

\author[0000-0001-9269-8167]{Richard D'Souza}
\affiliation{Vatican Observatory, Specola Vaticana, I-00120, Vatican City State, Italy}

\author[0000-0001-6380-010X]{Jeremy Bailin}
\affiliation{Department of Physics and Astronomy, University of Alabama, Box 870324, Tuscaloosa, AL 35487-0324, USA}

\author[0000-0001-6982-4081]{Roelof S.\ de Jong}
\affiliation{Leibniz-Institut f\"{u}r Astrophysik Potsdam (AIP), An der Sternwarte 16, D-14482 Potsdam, Germany}

\author[0000-0002-2502-0070]{In Sung Jang}
\affiliation{Department of Astronomy and Astrophysics, University of Chicago, Chicago, IL 60637, USA}

\author[0000-0002-0558-0521]{Colin T.\ Slater}
\affiliation{Department of Astronomy, University of Washington, Box 351580, Seattle, WA 98195-1580, USA}

\begin{abstract}
Mergers of and interactions between galaxies imprint a wide diversity of morphological, dynamical, and chemical characteristics in stellar halos and tidal streams. Measuring these characteristics elucidates aspects of the progenitors of the galaxies we observe today. The M81 group is the perfect galaxy group to understand the past, present, and future of a group of galaxies in the process of merging. Here we measure the end of star formation $(t_{90})$ and metallicity $([M/H])$ of the stellar halo of M82 and the eastern tidal stream of NGC 3077 to: 1) test the idea that M82 possesses a genuine stellar halo, formed before any interaction with M81, 2) determine if NGC 3077's tidal disruption is related to the star formation history in its tails, and 3) create a timeline of the assembly history of the central trio in the M81 group. We argue that M82 possesses a genuine, metal poor ([M/H] $\sim$ $-$1.62 dex) stellar halo, formed from the merger of a small satellite galaxy roughly 6.6 Gyr ago. We also find that the stars present in NGC 3077's tails formed before tidal disruption with M81, and possesses a roughly uniform metallicity as shown in \protect\cite{Okamoto2023} implying that NGC 3077's progenitor had significant population gradients. Finally, we present a timeline of the central trio's merger/interaction history.
\end{abstract}

\section{Introduction}
\label{sec:intro}
 
The interactions between and eventual merging of galaxies fundamentally changes their structure and composition \citep{Toomre&Toomre1972,White&Rees1978,Bullock2005}. Many factors shape the effects of these dramatic events, but the mass of the interacting/merging galaxies as well as the time since the event are two of the most critical. Strong interactions tidally disrupt the less massive of the interacting pair, affecting the star formation activity and altering stellar population gradients along the tidal tails \citep[e.g.][]{Taibi2022}. Large mergers disrupt discs and bring in material that fuels star formation and black hole growth \citep{Toth1992,DiMatteo2005,Sotilli-Ramos2022}, and while less common than mergers at early times, recent mergers (within the last $\sim$9 Gyr) have the most substantial consequences for galactic morphology \citep{Sotilli-Ramos2022}. 

Our main observational probes of this hierarchical growth are the accreted stars that are tidally stripped from the satellite galaxies. The early stages of tidal disruption result in tidal streams: measurements of their properties are sensitive to the orbit and progenitor properties (e.g., \citealt{Fardal2007}, \citealt{Law2010}). This material eventually phase mixes into a stellar halo in which the contribution from individual satellites is more difficult to distinguish. The vast majority of a stellar halo's mass is accreted in a single event, or a ``most dominant" merger \citep{Deason2016, DsouzanBell2018, Monachesi2019}. This makes the mass, metallicity, and star formation history (SFH) of the stellar populations deposited in tidal streams and stellar halos the most effective probe into the history of a galaxy, existing as a ``fossil record" of a galaxy's interaction and merger history. 

Tidal streams and accreted stars have illuminated the hierarchical growth history of the Milky Way and M31. The Milky Way has many known stellar streams, from both disrupting satellite galaxies and star clusters (e.g., \citealt{Shipp2018}). The Milky Way's most massive recent accretions are the Magellanic Clouds; orbital modeling and the {\sc Hi}-rich Magellanic Stream \citep{Putman98,Nidever2010} suggests that they fell in as a group, along with a number of smaller dwarf satellites \citep{Kal13,Patel2020}. The ongoing disruption of the Sagittarius Dwarf Spheroidal has given rise to huge tidal streams that wrap entirely around the Milky Way \citep{Ibata97,Majewski2003}. In addition to providing information about the potential of the Milky Way \citep{Law2010}, stellar population gradients along the stream inform our understanding of the progenitor galaxy and the tidal disruption process \citep[e.g.][]{Keller2010,Hayes2020,Cunningham2024}. 
Yet, the bulk of our stellar halo is debris from the Milky Way's most massive ancient accretion, the Gaia Sausage-Enceladus (GSE) event \citep[e.g.][]{Belokurov2018,Haywood2018,Helmi2018}; this event happened around $\sim9$\,Gyr ago, as evidenced by a shut-off in star formation of GSE stars at that time \citep{Gallart2019}. 

 
In contrast, the tidal streams around and stellar halo of M31 are much more massive and prominent than those of the Milky Way. The Giant Stellar Stream has an order of magnitude more mass than the Sagittarius Stream, and is associated with several shells discernible in velocity--position space (e.g., \citealt{Ibata01,Fardal2007,Dey2023}).  The Giant Stream has a considerable population of intermediate-aged stars \citep{Brown2006}, and possesses significant population gradients \citep[e.g.][]{Escala2021}. Several models for the stream have been proposed --- all models agree that the stream was produced relatively recently in the last 2--3\,Gyr, but a range of stellar masses from $M_* = 10^9 - 2\times 10^{10} M_\odot$ have been proposed \citep{Fardal2007,Hammer2018,DsouzanBell2018,Milo2022}. M33 is M31's most massive satellite, and appears to have a tidally-distorted stellar and gas envelope \citep{McConnachie09}, and possibly a stellar halo \citep{Gilbert2022,Smercina2023}.  M31's stellar halo is massive and metal-rich (10-20$\times$ the Milky Way halo's mass, and $\sim5\times$ its metallicity; \citealt{Ibata2014}), and has a significant intermediate-aged population \citep{Brown2006,Brown2008}. These halo properties are thought to reflect a dominant merger happening $\sim 2-3$\,Gyr ago \citep{DsouzanBell2018}, which is thought to have thickened M31's stellar disk and triggered a galaxy-wide starburst \citep{Williams2017,Hammer2018}. 

\begin{figure*}[bht]
    \centering    \includegraphics[width=\linewidth]{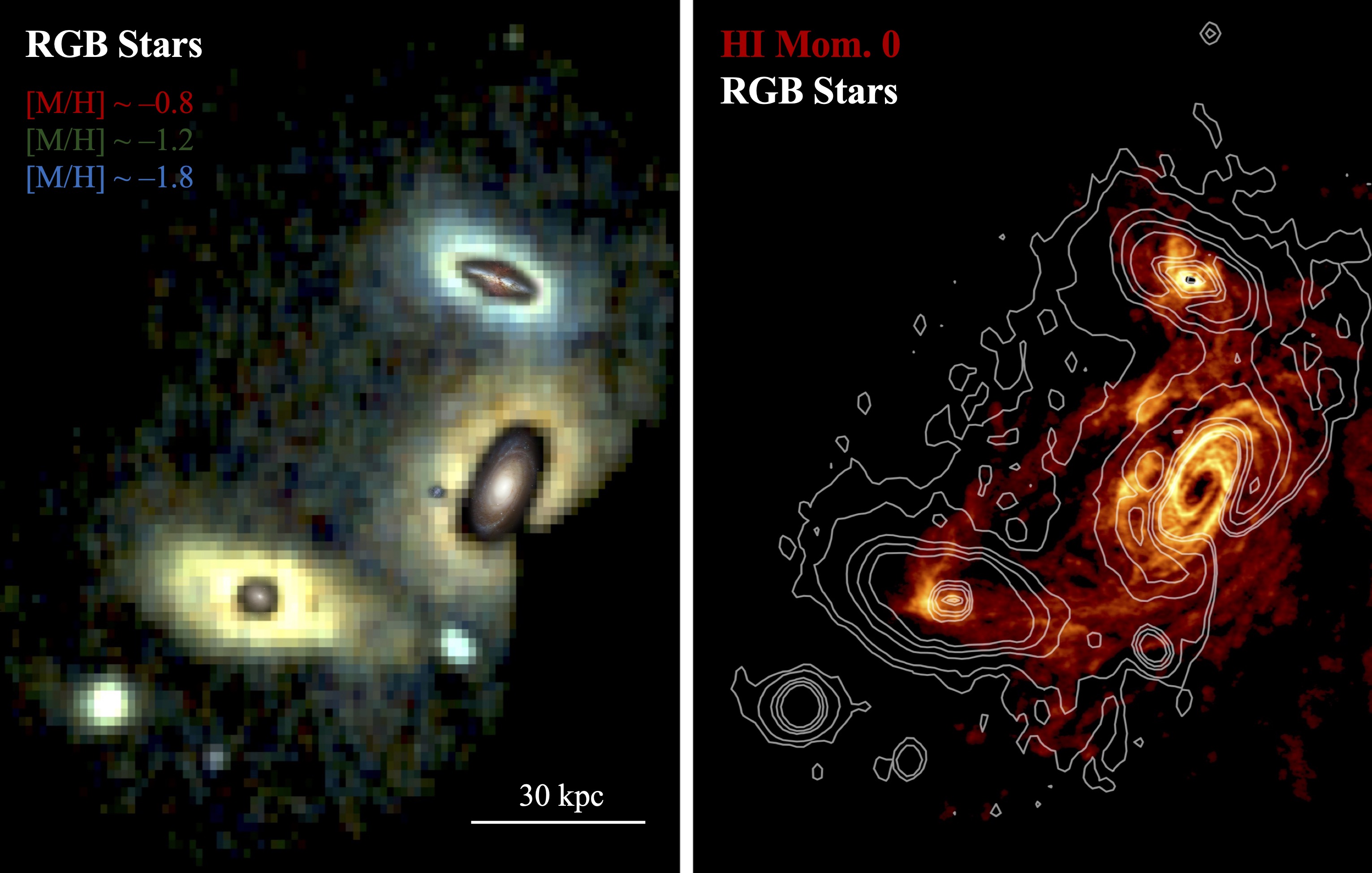}
    \caption{An introduction to the M81 Group environment and the different views of its triple interaction. \textit{Left}: A three color image of the M81 group taken from \cite{Smercina2020} showing the density of individual RGB stars in the outskirts of these three galaxies. The low metallicity of M82's halo and the tidally liberated material enveloping NGC 3077 are clearly visible. \textit{Right}: A map of \textsc{h\,i} \citep{deBlok2018} in the M81 Group, shown in the red colormap, on the same scale as the stellar halo. The contours of the total RGB stellar density are overlaid in white for direct comparison (6, 10, 30, 50, 80 RGB/arcmin$^2$ calculated on a map smoothed with a 0.65$\sigma$ Gaussian kernel). Tidal streamers of diffuse \textsc{h\,i} are visible over nearly the entire stellar tidal field.}
    \label{fig:bigmap}
\end{figure*}

As ground-based detection limits get deeper, the mass, metallicity, and star formation history of more distant global stellar halo and tidal stream populations are readily available for observation. The M81 group in particular is an ideal candidate for studying how galaxies change during a massive merger (Figure\ \ref{fig:bigmap}). Each of the central galaxies, M81, M82, and NGC 3077, have garnered the interest of researchers for many years, but there is a growing fascination with the rich interaction history of these three galaxies. Widespread \textsc{h\,i} tidal debris and unbound stellar material are direct evidence of recent interactions \citep[e.g.][also Figure\ \ref{fig:bigmap}]{Yun1999,deBlok2018,Smercina2020}. Some of the denser regions of gas are actively forming stars \citep{Okamoto2015,Okamoto2019}. \cite{Smercina2020} predict that the central trio of galaxies in this group will merge in the next 2 Gyr, meaning this group of galaxies is in a very transient life stage. Studying the timescales associated with these galaxies' interaction history therefore gives a unique look at the past, present, and future of a massive halo in formation. 

Past works on this central set of galaxies (M81, M82, and NGC 3077) found the distribution of different evolutionary stages of stars \citep{Okamoto2015}, characteristics of young stellar systems \citep{Okamoto2019}, and new dwarf galaxy candidates in nearby regions \citep{Bell2022}. \cite{Monachesi2013,Monachesi2016a} found the halo of M81 to be relatively metal poor ([M/H] $\sim -1.2$ dex). \cite{Smercina2020} recently uncovered the interaction history of M81, showing how it had a relatively quiet past, only accreting at most an SMC mass galaxy over the course of its life. \cite{Durrell2010} found this halo to be $9 \pm 2$ Gyr old by comparing the Red Giant Branch to theoretical isochrones, giving a ``when" to this SMC-sized merger, fully characterizing this stellar halo and the life history of M81. However, few studies have been done on the extended structures of the other two galaxies in the central trio. 

M82, the second-most massive in the group, is a starburst galaxy about half the mass of the Milky Way ($M_* \sim 2.8{\times}10^{10}\,M_{\odot}$; \citealt{querejeta2015}). It is currently interacting with M81, leading to uncertainty if its stellar halo is ``genuine", that is, formed from a merger event before its infall to the group. However, \cite{Smercina2020} show how M82's outskirts are metal poor and appear to be not significantly tidally disrupted, leading us to believe that this structure is indeed a halo that formed long before any interaction with M81. \cite{Smercina2020} do not present a well-constrained metallicity of M82's halo, motivating our measurement of it in this work. 

The third and smallest of this trio, NGC 3077, is an irregular dwarf galaxy similar in mass to the LMC ($M_* \sim 2.3{\times}10^{9}\,M_{\odot}$; \citealt{querejeta2015}). \cite{Okamoto2015,Okamoto2023} revealed that NGC 3077 exhibits an S-shaped structure, a characteristic of tidal disruption. \cite{Okamoto2023} also report that the metallicity of the tidal tails is the same as that of the outer edge of the main body of the galaxy ($\sim -1.4$ dex), showing that the tidal debris most likely originated from the outer envelope of NGC 3077. 

These works have characterized parts of the properties of M82 and NGC 3077's extended structures, but their star formation histories are unknown and metallicities have been sparsely measured in the past. To learn the full history of this group, important questions still need to be answered: When did star formation stop in the outskirts of these galaxies? Is this shutoff connected to their metallicities or dynamical histories? How do these star formation timescales compare with the assembly and evolutionary timescales of the rest of the group?

To answer these questions, measurements of stellar mass, metallicity, and age need to be made of the outskirts of M82 and NGC 3077. The first two of these quantities are relatively easy to obtain observationally and have been found to some degree in the past, but finding the ages of stars in these diffuse structures is a non-trivial task. Traditional methods require fitting stellar isochrones to color magnitude diagrams (CMDs) to infer age, but the majority of stars in these distant, low surface brightness features are below the detection limits of current instruments. The methods in \cite{Harmsen2023} circumvent this problem by establishing a relationship between the two of the brightest stellar evolutionary stages, being the Asymptotic Giant Branch (AGB) and the upper Red Giant Branch (hereafter TRGB), and the lookback time at which the majority of star formation stopped. While crude, this is the only way to measure the ages of stars in the outskirts of distant galaxies. Making these measurements and combining them with past work on this group will allow us to broadly characterize the past, present, and future of this group's assembly and evolution. 

Section \ref{sec:obs_n_data} outlines the our observations and data reduction techniques, as well as star galaxy separation and completeness corrections. Section \ref{sec:selections} describes the spatial and color-magnitude selections for our measurements, the methods to calculate t$_{90}$, and how we measured the metallicity of our spatial selections. Section \ref{sec:results} describes the derived t$_{90}$ values of the selections and our metallicity measurements, and Section \ref{sec:interp} provides a timeline of the M81 group's evolution as well as interpretations of our measurements of M82 and NGC 3077. We conclude in Section \ref{sec:conclusions}.

\section{Observations, Data Reduction, and Completeness Correction}
\label{sec:obs_n_data}

\begin{figure}[bht]
    \centering
    \includegraphics[width=\linewidth]{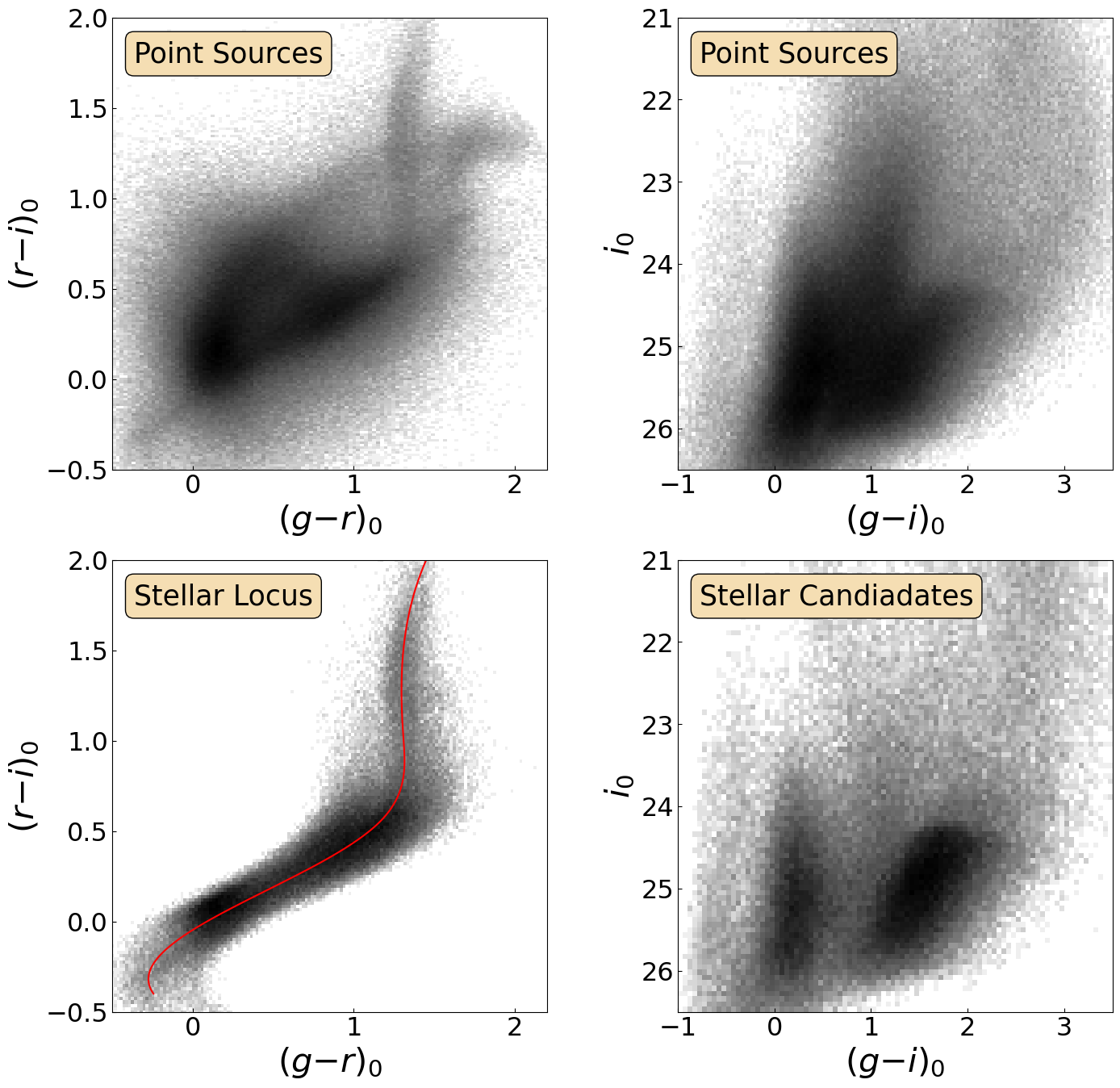}
    \caption{Panels showing total unresolved point sources, (upper) and stellar candidates that lie along the stellar locus (within 0.2 mag) (lower). The difference between the point source CMD (upper right) and the stellar candidates CMD (lower right) is clear, with a large number of background sources eliminated. A significant number of background galaxies remain (\emph{g}$\minus$\emph{i} $\sim$ 0) that need to be removed through background subtraction.}
    \label{fig:stargal}
\end{figure}

The \emph{Subaru Telescope HSC} was used for two nights of observation through the Subaru-Gemini exchange program on March 26th and 27th, 2015 (PI: Bell, 2015A-0281). The outer regions of M81, M82 and NGC 3077 were observed in each the \emph{g}-, \emph{r}-, and \emph{i}-bands using two pointings. The number of exposures and integration times can be found in Table 1 of \cite{Smercina2020}. 

The data were reduced with the HSC pipeline \citep{Bosch2018} that uses the Pan-STARRS1 catalogue \citep{Magnier2013} to perform photometric and astrometric calibration, as well as converting the photometry to the SDSS photometric system. The specific data reduction that was used performed background subtraction with a 32-pixel mesh that is optimized for point sources to remove the majority of diffuse light. The \emph{i}-band was used to provide reference positions and image sizes to force photometry in the \emph{g}- and \emph{r}-bands, as it is the deepest and best-resolved. 

The galactic extinction was corrected following \cite{Finkbeiner2011}. They estimate the E(B-V) near the M81 group to be $\sim$0.1. The extinction-corrected point source detection limits were fairly uniform across the two fields, at \emph{g} = 27.0, \emph{r} = 26.5, and \emph{i} = 26.2, all measured at 5$\sigma$. Stable seeing during the two nights of observation led to point-source sizes near 0.7" - 0.8" down to the detection limits, resulting in the observation of nearly 900,000 sources across the two fields. For further discussion of the pipeline, see \cite{Bosch2018}. 

Due to the distance of the M81 group and the diffuse nature of stellar halos, stellar sources were greatly outnumbered by background galaxies. To combat this, extensive star galaxy separation was done to ensure the purity of our sample. The first approach was to consider the morphology of sources unresolved at the resolution limit of our observations. Limiting the FWHM of sources to $\leq$0.9" removed the majority of extended background sources in our observations. 

As our study relies on the number of stars in a particular evolutionary stage contained in a selection of a diffuse stellar halo, more stringent requirements were needed to limit background count contamination. The next approach to reduce contamination was to require that true stellar sources lie within 0.2 mag (taking into account photometric error) of the ``stellar locus" in the \emph{g} $\minus$ \emph{r} vs. \emph{r} $\minus$ \emph{i} color-color space \citep[following][]{Covey2007,High2009,Smercina2017}. The stellar locus is depicted in Figure \ref{fig:stargal} (lower left) as a red line. Sources that passed both of our requirements are shown in the lower right of Figure \ref{fig:stargal}, showing $\sim$160,000 sources compared to the initial $\sim$900,000. 

While many background contaminants were rejected through the methods outlined above, a significant number of compact background galaxies remain. To remove these from our spatial samples, sections of empty sky of equal area to the sections of interest were subtracted from said sections of interest (see Section \ref{sec:CMDselections} and Figure \ref{fig:spatial3} for the spatial selections and their respective reference selections). Additionally, our survey becomes crowded and therefore incomplete \citep{Smercina2020} toward the centers of these three galaxies, and these dense regions were masked out and do not appear in the images of the fields that are included in this paper.

\begin{figure}[thb]
    \centering
    \includegraphics[width=\linewidth]{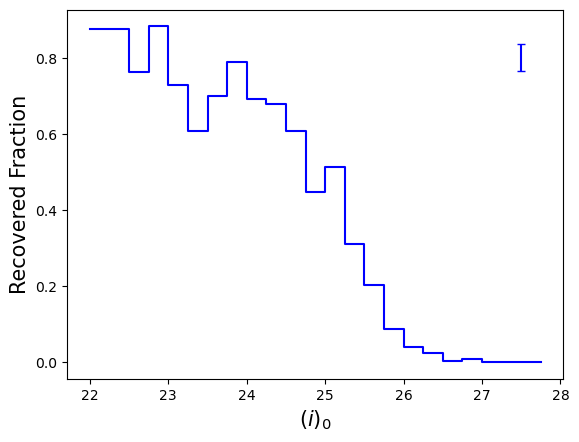}
        \caption{The fraction of HST point sources recovered to be stars using our morphological and color-color selections at a given \emph{i}-band magnitude. The typical uncertainty per bin is show in the upper right, with a value of $\pm 3.5\%$. }
    \label{fig:comp}
\end{figure}

Our data was compared to existing GHOSTS survey data in the halo of M81 \citep{Radburn-Smith2011,Monachesi2016a}, and incompleteness curves (Figure \ref{fig:comp}) and corrections were made to the counts found in uncrowded areas. For both M82 and NGC 3077, stellar counts were divided by the average recovered fraction of sources at a given \emph{i}-band magnitude within the AGB and TRGB selections. In the halo of NGC 3077, a galactocentric distance was found that had the same stellar surface density as the inner radius of our selection of the halo of M82 to ensure crowding was not an issue in our corrections. We prefer to use empirical completeness corrections from real HST data in order to circumvent systematic error caused by imperfect PSF and color-color modeling for artificial stars --- in practice, adoption of artificial star-derived completeness corrections does not affect our inferences to within our error bars. In Fig.\ \ref{fig:comp}, we show the typical uncertainty in completeness, determined from assessing completeness in separate areas of the field. Folding in these completeness uncertainties into our analysis, we find that these lead to 0.02 dex variations in log$_{10}$(N$_{*,AGB}$/N$_{*,TRGB}$), corresponding to 0.3 Gyr in $t_{90}$, well within our random and systematic uncertainties.

\section{Selections and Measurements}
\label{sec:selections}
\subsection{Color-Magnitude Diagram and Spatial Selections}
\label{sec:CMDselections}

\begin{figure}[th]
    \centering
    \includegraphics[width=\linewidth]{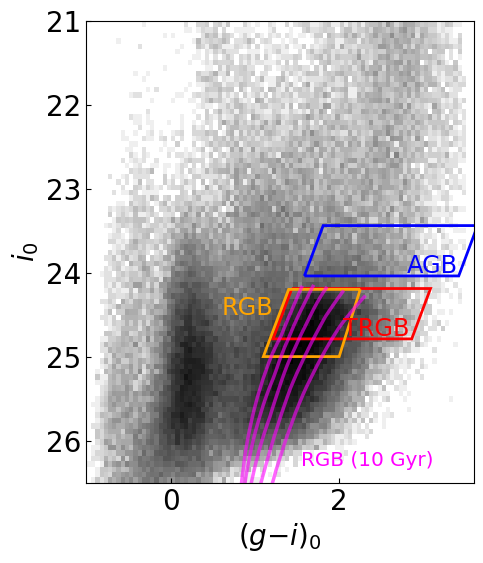}
    \caption{The CMD of our final stellar candidates. Background contaminates can be seen in the dense region between \emph{i} $\sim$ 24 and \emph{i} $\sim$ 26 and \emph{g} $\minus$ \emph{i} $\sim$ 0. 10 Gyr RGB isochrones are included to understand the location of this distinct stellar evolutionary stage. The metallicities of these isochrones range from [M/H] = -2.0 to -0.8, with steps of -0.3 dex.}
    \label{fig:cmd1}
\end{figure}

Our final star-galaxy separated stellar candidates appear in Figure \ref{fig:cmd1}. The majority of the sources that appear there are at the distance of the M81 group, although many foreground Milky Way (MW) stars and background sources are still present. We highlight the selection boxes used to AGB stars and TRGB stars in Figure \ref{fig:cmd1}. These selections were defined and calibrated in \cite{Harmsen2023} around a placement of the tip of the Red Giant Branch (23.72 in F814W at the distance of M81 following \citealt{Radburn-Smith2011}) using Hubble Space Telescope (HST) magnitudes, so a conversion was done between the HST F606W/F814W and SDSS \emph{g}/\emph{i}-band using isochrones generated for both magnitude systems (PARSEC; \citealt{Bressan2012}). We also show the portion of the Red Giant Branch used to calculate metallicity; we use the \emph{i} $\leq$ 25 up to the tip of the Red Giant Branch for a more complete sample. 10 Gyr RGB isochrones with metallicity ranging from [M/H] = -2.0 to -0.8 were included (Figure \ref{fig:cmd1}) to better understand the location of this stellar evolutionary stage. Due to contamination from helium burning stars on the blue side of the AGB and TRGB selections, they were shifted by 0.4 in color to ensure a pure sample of these stars. This shift potentially removed a non-trivial number of AGB and TRGB stars, but we chose to prioritize the purity of the selection over completeness. (See Appendix \ref{appendix_heb})

\begin{figure*}[tbh]
    \centering
    \includegraphics[width=\linewidth]{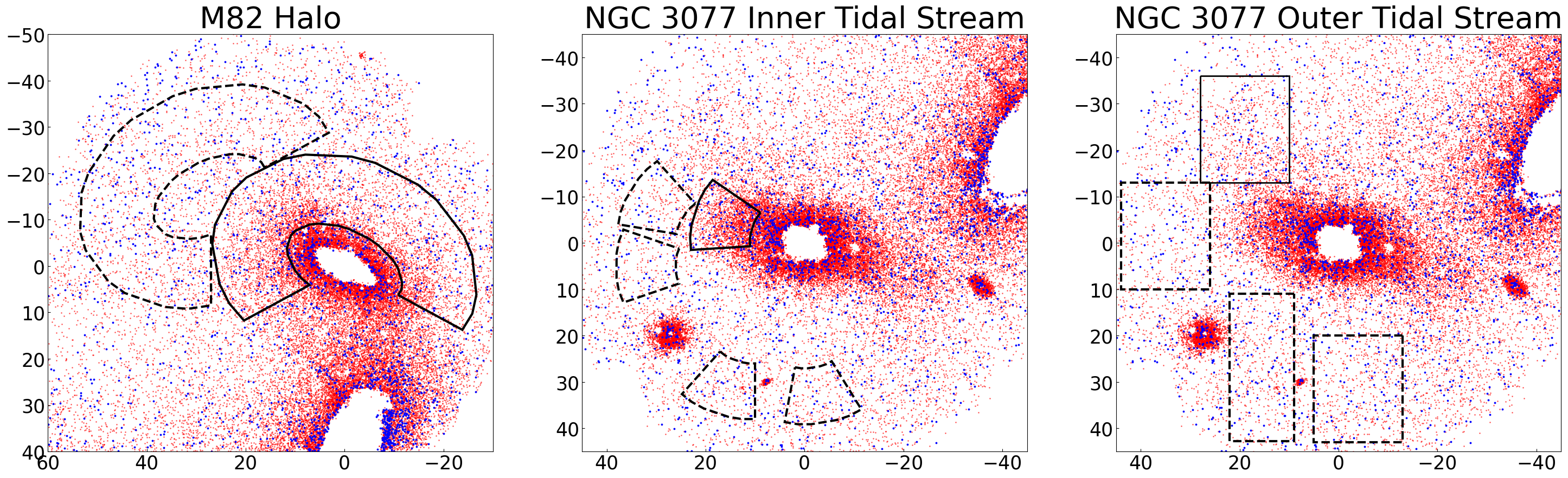}
    \caption{The selections (solid lines) described in Section \ref{sec:selections} of the halo of M82 (left) and eastern tidal stream (center and right) of NGC 3077. Background selections are shown in dashed lines. The axes are in units of kpc, with the origin centered on M82 in the left and NGC 3077 in the center and right. Red Giant Branch stars are shown in red points, and Asymptotic Giant Branch stars in blue. Large tidal bridges can be seen connecting these two galaxies to M81.}
    \label{fig:spatial3}
\end{figure*}

Spatial selections in the outskirts of both M82 and NGC 3077 were chosen to exclude regions potentially contaminated by the presence of M81 and the surrounding tidal bridges (Figure \ref{fig:spatial3} to see selections). \cite{Harmsen2023} points out that the M81 group is a prime example of a system where \emph{in-situ} halo star formation is actively occurring, requiring us to avoid selecting regions where known star formation is taking place. The selection of M82's halo ranges between 25 and 55 kpc along the major axis and 15 to 45 kpc along the minor axis, tracing out a 240 degree fraction of an annulus just outside of the outer disc/inner halo. We omit the southern 120 degree section to avoid contamination from the tidal bridges connecting M82 with M81, as we are only interested in the stellar populations that were present before the most recent interaction between these two galaxies. 

We follow \cite{Okamoto2015,Okamoto2023,Smercina2020} in the attribution of NGC 3077's entire envelope to tidal tails formed from its main body. Although this diffuse outer structure is not a stellar halo, analyzing these tails will help to answer our questions about the origin of the stars composing these structures and any connection between the ages and metallicities with dynamical history. The stream to the southwest of NGC 3077 is known to have more recent star formation than the stream on the northeast, as well as is projected closer to M81. For these reasons, the tidal stream on the northeast was analysed in this work. NGC 3077's inner tidal stream selection is a 70 degree section on the eastern side of the galaxy, stretching from 22 to 46 kpc away from the galactic center. We also elected to analyze one side of NGC 3077's faint outer tidal features. This selection is 18 kpc by 23 kpc, as first shown in \cite{Okamoto2023}. 

When possible, multiple background selections were used and their counts were averaged to improve the number statistics of each respective halo selection. Additionally, these galaxies and their outskirts reside near/within the larger stellar halo of M81. The choice of background selections to be close to or far from the stellar halo of M81 does not affect our measurements within our error bars. 

\vfill\null
\subsection{t$_{90}$ Calculation}
\label{sec:t90_calc}

To determine the ages of the stellar populations in the outskirts of M82 and NGC 3077, we follow the method outlined in \cite{Harmsen2023}. The log$_{10}$ of the ratio of the number of AGB stars (N$_{*,AGB}$) to TRGB stars (N$_{*,TRGB}$) in the spatial selections were found and converted into t$_{90}$ using the below equation: 
\begin{equation}
t_{90,expect}=t_{90,-1}+\alpha(\log_{10}(N_{*,AGB}/N_{*,TRGB})+1)
\end{equation} 
where the intercept $t_{90,-1} = 4.4 \pm 0.3$ is the expected $t_{90}$ at $log_{10}(N_{*,AGB}/N_{*,TRGB}) = -1$ and $\alpha = -11.3 \pm 1.6$ is the slope of the relation. This relation has an intrinsic scatter of $\sigma_t = 1.45 \pm 0.18$ Gyr. We take this intrinsic scatter and the error on each of these parameters in quadrature with the counting error from our selections. 

The stellar counts, ratios, and t$_{90}$ for each selection are found in Section \ref{sec:results}, Table \ref{Tab:table}. The halo of M82  showed no significant spatial or radial variation in t$_{90}$ within the selected annulus, so we chose to present the global value. The spatial selections of NGC 3077 were sparse enough to require each selection being treated as one region, as opposed to breaking them up to create a more fine profile.

\subsection{[M/H] Measurements}
\label{sec:MH_method}

\begin{figure}[hbt]
    \centering
    \includegraphics[width=\linewidth]{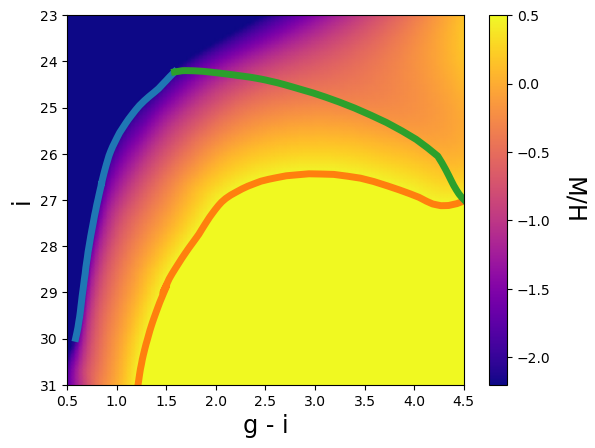}
    \caption{The interpolated metallicity map used to determine the [M/H] of our spatial selections. The left, blue line corresponds to the most metal poor isochrone used to determine metallicity ([M/H] = -2.15) and the right, orange line is the metal rich ([M/H] = 0.5). The green line connecting the two isochrones is the TRGB in the metallicity range used for our interpolation (-2.15 $<$ [M/H] $<$ 0.5).}
    \label{fig:MH_interp}
\end{figure}

To measure the metallicity of each spatial selection, we followed the methods outlined in \cite{Ogami2024}. We generated 54 isochrones from the PARSEC \citep{Bressan2012} isochrone suite with -2.15 $<$ [M/H] $<$ +0.5 at 0.05 dex intervals and an assumed an age of 10 Gyr. We construct a model of [M/H] as a function of \emph{i} magnitude and \emph{g - i} color using the \texttt{Scipy} radial basis function interpolation scheme (Figure \ref{fig:MH_interp}). The color and magnitude coordinates of each point source identified as both an RGB star based on the orange RGB selection in Figure \ref{fig:cmd1} and residing within our spatial selections was used to estimate the mean photometric metallicity. We do the same for the background selections. Both the regions of interest and background [M/H] distributions were binned and the difference between the two was found. The median [M/H] value and standard deviation of the resulting distributions are reported in this work in Table \ref{Tab:table}. Our choice of presenting the median versus the geometric mean does not affect our measurements outside of the error bars. Formal errors are very small ($<0.04$\,dex); systematic uncertainties dominate. We choose to adopt a simple systematic uncertainty reflecting the impact of stellar population age, by repeating our [M/H] calculation with ages of 6\,Gyr and 13\,Gyr respectively, resulting in variations in median metallicity of $\pm 0.20$ dex. We adopt  $\pm 0.20$ dex as an approximate median metallicity uncertainty in what follows. 

\begin{figure*}[thb]
    \centering
    \includegraphics[width=\linewidth]{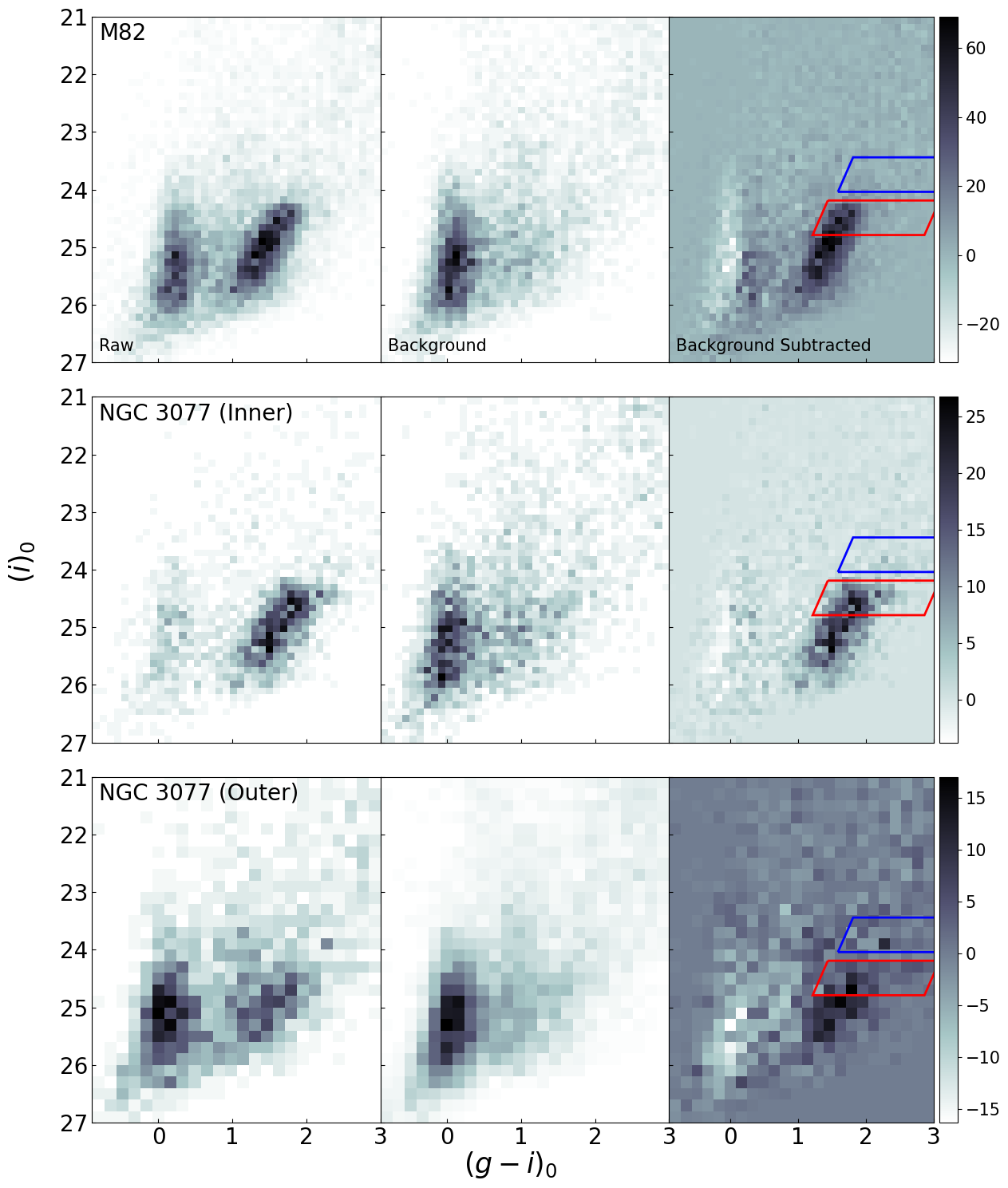}
    \caption{Raw (left), background (center), and background subtracted (right) CMDs for each of the selections shown in Figure \ref{fig:spatial3}, with the AGB and TRGB selections shown in the rightmost panel of each. The color bars show the stars/bin in each of the background subtracted CMDs. Counts in these selections and their corresponding ratios and t$_{90}$ are shown in Table \ref{Tab:table}}
    \label{fig:CMD_all}
\end{figure*}


\section{Results}
\label{sec:results}

The spatial distribution of stars that passed our stellar candidate criteria and selections for each region of both M82 and NGC 3077 can be seen in Figure \ref{fig:spatial3}. The extended structures of these galaxies are clearly visible, mainly dominated by RGB stars in red but with a fairly uniform distribution of AGB stars shown in blue. The tidal bridges connecting these galaxies with M81 are also visible. 

The CMD of M82's raw selection (Figure \ref{fig:spatial3}, left) as well as the background and background-subtracted CMDs are shown in Figure \ref{fig:CMD_all}. There is a fairly distinct Red Giant Branch, as well as a number of remaining background contaminants at (\emph{g} $\minus$ \emph{i}) $\sim$ 0. Relatively few AGB stars fall into our selection box ($\sim120$), as compared to the number in the TRGB ($\sim$2000). The log$_{10}$(N$_{*,AGB}$/N$_{*,TRGB}$) is $-$1.19 $\pm$ 0.06, leading to a t$_{90}$ of 6.6 $\pm$ 2.6 Gyr. Our [M/H] measurement of the spatial selection resulted in a value of $-$1.62 $\pm$ 0.20 dex. This qualitatively matches the metal poor appearance of M82 in Figures 12 and 14 of \cite{Smercina2020} who show multiple metallicity color-maps of the triplet and surrounding structures. 

The t$_{90}$ of both of the tidal stream selections for NGC 3077 can also be seen in Table \ref{Tab:table}. The CMDs of the selections shown in Figure \ref{fig:spatial3} (center and right) appear in Figure \ref{fig:CMD_all} (background-subtracted on the right). These CMDs show a well defined Red Giant Branch and a few AGB stars, with little background contamination. The log$_{10}$(N$_{*,AGB}$/N$_{*,TRGB}$) are $-$1.11 $\pm$ 0.06 for the inner selection and -0.93 $\pm$ 0.23 for the outer selection. The t$_{90}$ of the inner selection was 5.7 $\pm$ 2.4 Gyr which was moderately older than the outer selection, having a t$_{90}$ of 3.6 $\pm$ 3.3 Gyr. We measure a inner tidal stream [M/H] of $-1.43 \pm 0.20$ and an outer stream [M/H] of $-1.56 \pm 0.20$. These match what \cite{Okamoto2023} find for these sections within error ($\sim -1.35$ to $-1.5$ dex across both sections).

\begin{deluxetable*}{ccccccc}[bth]
\tablecaption{
    \textnormal{
    A table containing the AGB counts, TRGB counts, log$_{10}$ of the ratio of the counts, the calculated t$_{90}$ and metallicity measurements of each spatial selection. The errors on the counts and ratios are a combination of the Poisson noise of each selection and the standard deviation of the counts in background fields when multiple background fields were used. The error on the t$_{90}$ are a combination of these errors with the error on the fit parameters given in \cite{Harmsen2023}. Errors in metallicity are the systematic uncertainty described in Section 
    \ref{sec:MH_method}. Different background placements do not affect our measurements outside of the error bars. }
    \label{Tab:table}
}
\tablecolumns{2}
\setlength{\extrarowheight}{4pt}
\tablewidth{\linewidth}
\tabletypesize{\small}
\tablehead{
\colhead{Galaxy} & 
\colhead{N$_{*,AGB}$} & 
\colhead{N$_{*,TRGB}$} & 
\colhead{log$_{10}$(N$_{*,AGB}$/N$_{*,TRGB}$)} & 
\colhead{t$_{90}$ (Gyr)} & 
\colhead{[M/H]} & 
\colhead{$\sigma_{[M/H]}$}}
\startdata
\hline
M82 & 126 $\pm$ 21 & 1990 $\pm$ 50 & $-$1.20 $\pm$ 0.07 & 6.6 $\pm$ 2.6 & $-$1.62 $\pm$ 0.20 & 0.35\\ 
NGC 3077 (inner tidal stream) & 70 $\pm$ 9 & 908 $\pm$ 42 & $-$1.11 $\pm$ 0.06 & 5.7 $\pm$ 2.4 & $-$1.43 $\pm$ 0.20 & 0.37\\ 
NGC 3077 (outer tidal stream) & 16 $\pm$ 6 & 140 $\pm$ 54 & $-$0.93 $\pm$ 0.23 & 3.6 $\pm$ 3.3 & $-$1.56 $\pm$ 0.20 & 0.38\\
\hline
\enddata
\end{deluxetable*}

\section{Interpretation and Discussion}
\label{sec:interp}
Here we provide an inference of each galaxy's evolutionary history based on the t$_{90}$ and metallicity measurements presented above, as well as simulated analogs to contextualize these galaxies and their properties (Section \ref{sec:M82halo_interp} for M82 and Section \ref{sec:NGChalo/stream_interp} for NGC 3077). We also construct a timeline of the main trio in the group, composed of our measurements, those shown for the halo of M81 presented by \cite{Durrell2010}, and the starburst timescales of M82 and NGC 3077 from various literature sources (Section \ref{sec:timeline}). We end with a description of the limitations of our work and potential avenues to take similar work in the future (Section \ref{sec:limitations_future}).

\subsection{M82's Stellar Halo}
\label{sec:M82halo_interp}
\begin{figure*}[tbh]
    \centering
\includegraphics[width=\linewidth]{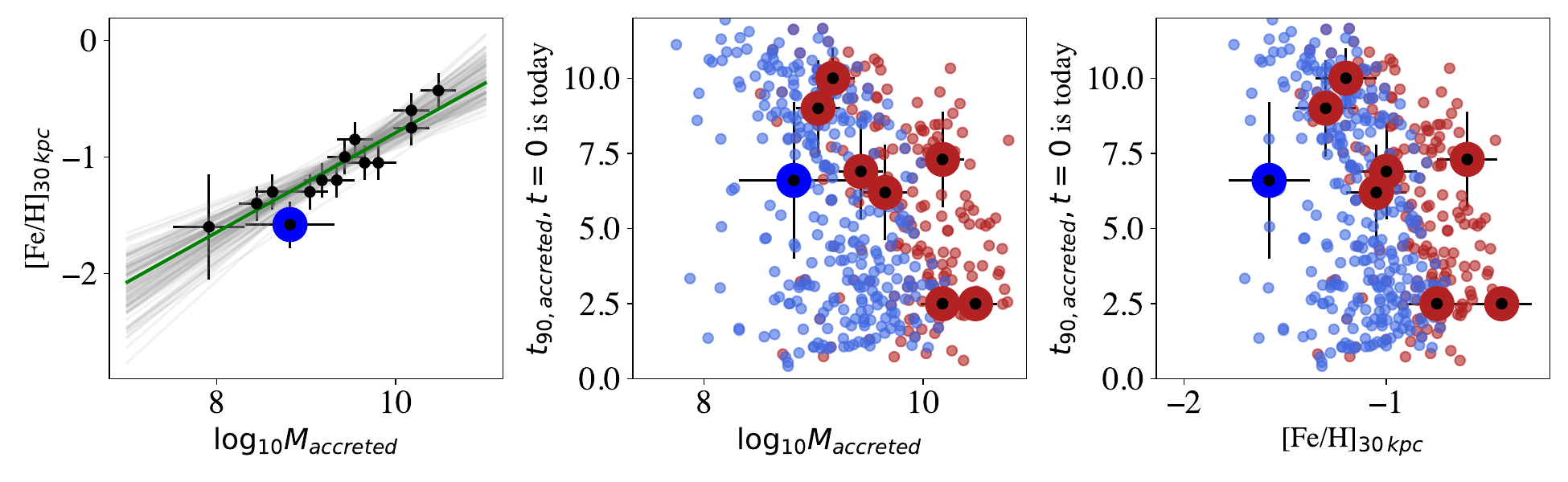}
    \caption{The stellar halo of M82 (large blue point) compared with stellar halos of MW mass galaxies reported in \cite{Harmsen2023} (large red points) and with simulated galaxies from TNG50 having similarly sized stellar halos to M82 (small blue points) and the MW (small red points). Left: stellar halo mass vs mean stellar metallicity of M82 and observed MW mass galaxies. Center: stellar halo mass vs t$_{90}$. Right: stellar halo mean metallicity vs t$_{90}$. The stellar halo mass of M82 was chosen to be three times that of the unbound mass reported in \cite{Smercina2020}, with the error ranging from one to five times the unbound mass. The TNG50 metallicities in the right panel were found using the green best-fit slope in the left panel.}
    \label{fig:sims}
\end{figure*}

This work gives new insights on M82's extended envelope of stars, that owing to being much poorer in metals than the disk has been previously argued to be a stellar halo \citep{Harmsen2017,Smercina2020}. From its CMD, we have estimated its metallicity to be $[M/H] = -1.62$\,dex, and on the basis of a relatively sparse AGB population estimate a $t_{90} \sim 6.6\pm2.6$\,Gyr. The mass of M82's stellar halo is considerably less certain.  We take a simplistic approach where we scale M82's mass outside the tidal radius $R_{tidal} \sim 10$\,kpc of $2.1\times 10^8 M_{\odot}$ \citep{Smercina2020}. Following \citet{Harmsen2017} and \citet{Bell2017}, we adopt a factor of three correction to correct a $>10$\,kpc mass to a total accreted mass of $M_{acc} \sim 6.3 \times 10^8 M_{\odot}$; we adopt a $\pm0.5$\,dex (factor of three) error in acknowledgement of the considerable uncertainty when dealing with a halo in a disrupting satellite galaxy. 

With these in hand, we can then ask how M82's stellar halo compares with those of other galaxies. In Figure \ref{fig:sims}, we show the stellar halo of M82 (large blue point) and the stellar halos of several other Local Volume galaxies (black points; left; large red points center and right panels). M82's stellar halo is consistent with the broad trend between stellar halo metallicity and stellar halo mass (left panel; data points from \citealt{Gozman2023}) --- recall that this relationship reflects that most of the stars in a stellar halo are provided by its most massive accreted satellite, which drives up both the metallicity (by the metallicity--mass relation of the accreted satellites) and the stellar mass \citep[e.g.,][]{Deason2016,Harmsen2017,DsouzanBell2018}. 

With a $t_{90}$ value of $\sim 6.6$\,Gyr, M82's halo has an age consistent with the median halo age (from \citealt{Harmsen2023}) of nearby galaxies. The comparison sample contains no halo ages for galaxies with comparable stellar mass; accordingly we compare with $t_{90}$ values for the accreted components of M82-mass (blue) and MW-mass (red) simulated galaxies from the TNG-50 simulation (central and right panels of Fig.\ \ref{fig:sims}). 
TNG-50 \citep{Pillepich2018,Pillepich2019}
simulates a large cosmological volume $\sim 50$\,Mpc on a side with $\sim$300\,pc resolution, enabling the analysis of the detailed properties of stellar haloes. The MW-mass galaxies (red) spanned 4$\times 10^{10} M_{\odot}$ to 15$\times 10^{10} M_{\odot}$; the M82-mass galaxies (blue) span $10^{10} M_{\odot}$ to 4$\times 10^{10} M_{\odot}$, having dark matter halo masses less than $10^{13} M_{\odot}$. We analyze `accreted' particles born out of the main progenitor branch subhalo. In common with \citep{Harmsen2023}, we calculate $t_{90}$ directly from the cumulative star formation history of the accreted particles\footnote{Choosing instead to focus on accreted particles with galactocentric radii $r>10$\,kpc (more like observed stellar halo stars) affects $t_{90}$ at the level of $\pm$1\,Gyr. We choose to focus on all accreted particles for simplicity and to compare directly with \citet{Harmsen2023}.}. Like the halos of more massive galaxies, the accreted stars of simulated M82-mass galaxies show a wide range of $t_{90}$ values, with a tendency towards younger $t_{90}$ values for more massive/metal-rich halos; the overall values and trend is offset towards rather lower accreted mass/metallicity. M82's halo is a little on the young side for its accreted mass (and particularly metallicity, owing to its rather low metallicity for its accreted mass).  

The implications for inferences about M82's accretion history are uncertain, and would benefit from both improved observational constraints and future simulations. Many papers have found that to a first approximation, the stellar halo mass and metallicity reflect the stellar mass and metallicity of a galaxy's most important accretion event \citep[e.g.,][]{Deason2016,Harmsen2017,DsouzanBell2018,Monachesi2019}. The interpretation of $t_{90}$ is less certain (as measurements of halo star formation histories are still rare), but is predicted to reflect the time of the most important accretions --- for example, the observed decrease in star formation in the halos of the MW \citep[e.g.,][]{Gallart2019} and M31 \citep[e.g.,][]{Brown2008} occurred at the time of the most dominant merger event (see also \citealt{DsouzanBell2018,Harmsen2023}). Taken then at face value, we would suggest that this halo formed from a merger event occurring $\sim$ 6.6 Gyr ago, with a metal-poor galaxy with roughly the mass of the SMC, similar to the merger partners of M64 \citep{Smercina2023} and M94 \citep{Gozman2023}.  

A possible complication to this picture is the low stellar mass, and seemingly low metallicity, of M82's stellar halo. Both \citet{DsouzanBell2018} and \citet{Monachesi2019} find that lower mass simulated stellar halos tend towards having contributions from several important satellites. Such halos would tend towards having lower metallicity for their stellar mass (as they are the combination of a few lower metallicity satellites rather than the debris of a more massive and metal-rich one; e.g., \citealt{Harmsen2017,DsouzanBell2018}), consistent with the (very uncertain) halo metallicity estimate $[M/H] = -1.62\pm0.20$. An alternate interpretation of low metallicity halos as signaling early accretion \citep[following][]{Harmsen2017} would be disfavored owing to M82's halo $t_{90} \sim 6.6$\,Gyr. Consequently, while we tentatively suggest that M82's most important accretion was $\sim 6.6$\,Gyr ago with a SMC-massed galaxy, it is possible that $t_{90} \sim 6.6$\,Gyr and M82's halo mass instead reflect the mean accretion time of a modest number of smaller satellites (e.g., a significant fraction of the SMC's mass). Simulations suggest that more information is available about accretion history with improved observations (e.g., structure, age structure/gradients, metallicity gradients, etc.; \citealt{Johnston2008,Cook2016,DsouzanBell2018,Monachesi2019,Wright2023}), highlighting the potential of upcoming deep JWST stellar observations (PI: Smercina). 

\subsection{NGC 3077's Tidal Streams}
\label{sec:NGChalo/stream_interp}

\begin{figure}[tbh]
    \centering
\includegraphics[width=\linewidth]{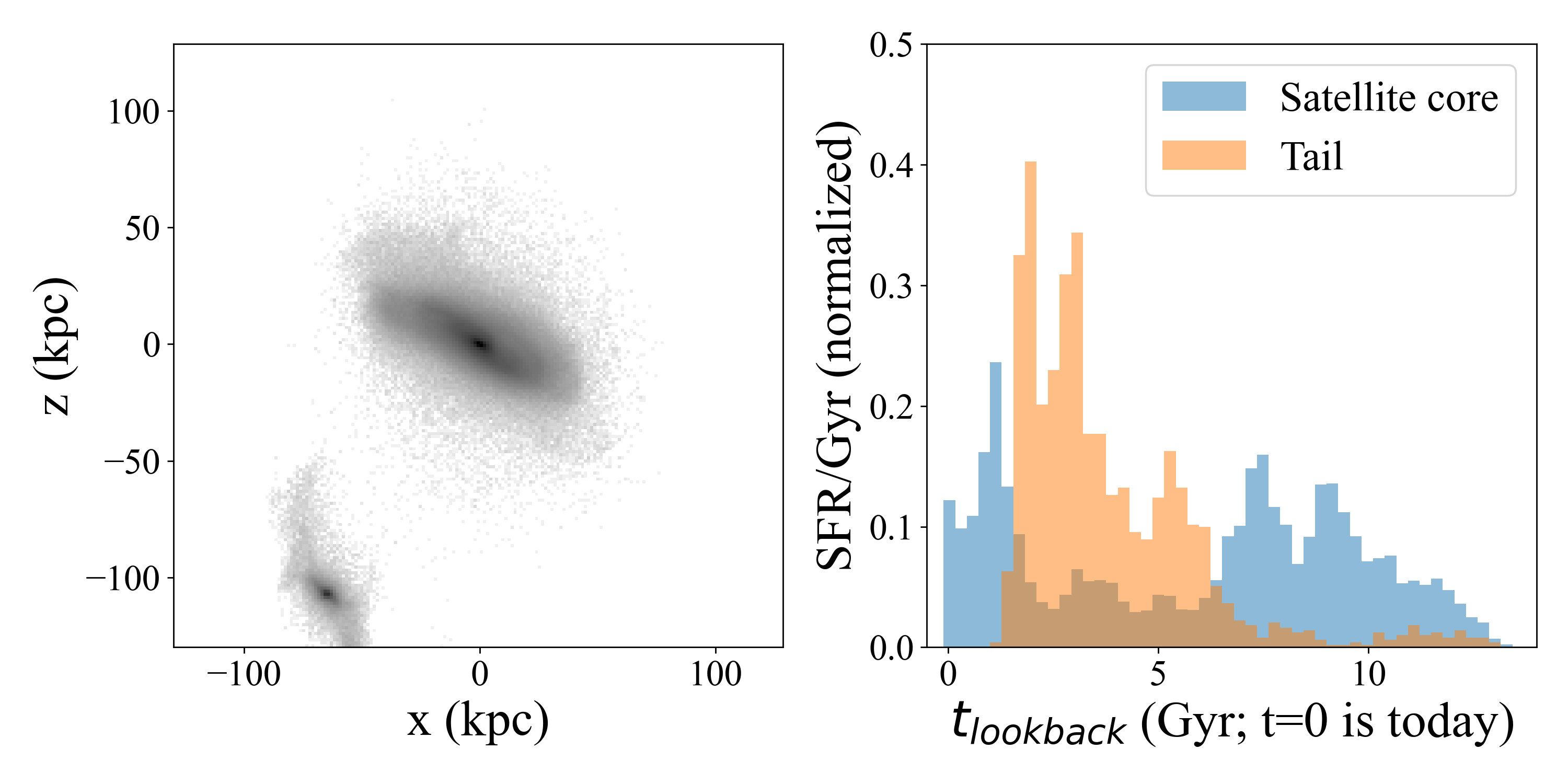}
\includegraphics[width=\linewidth]{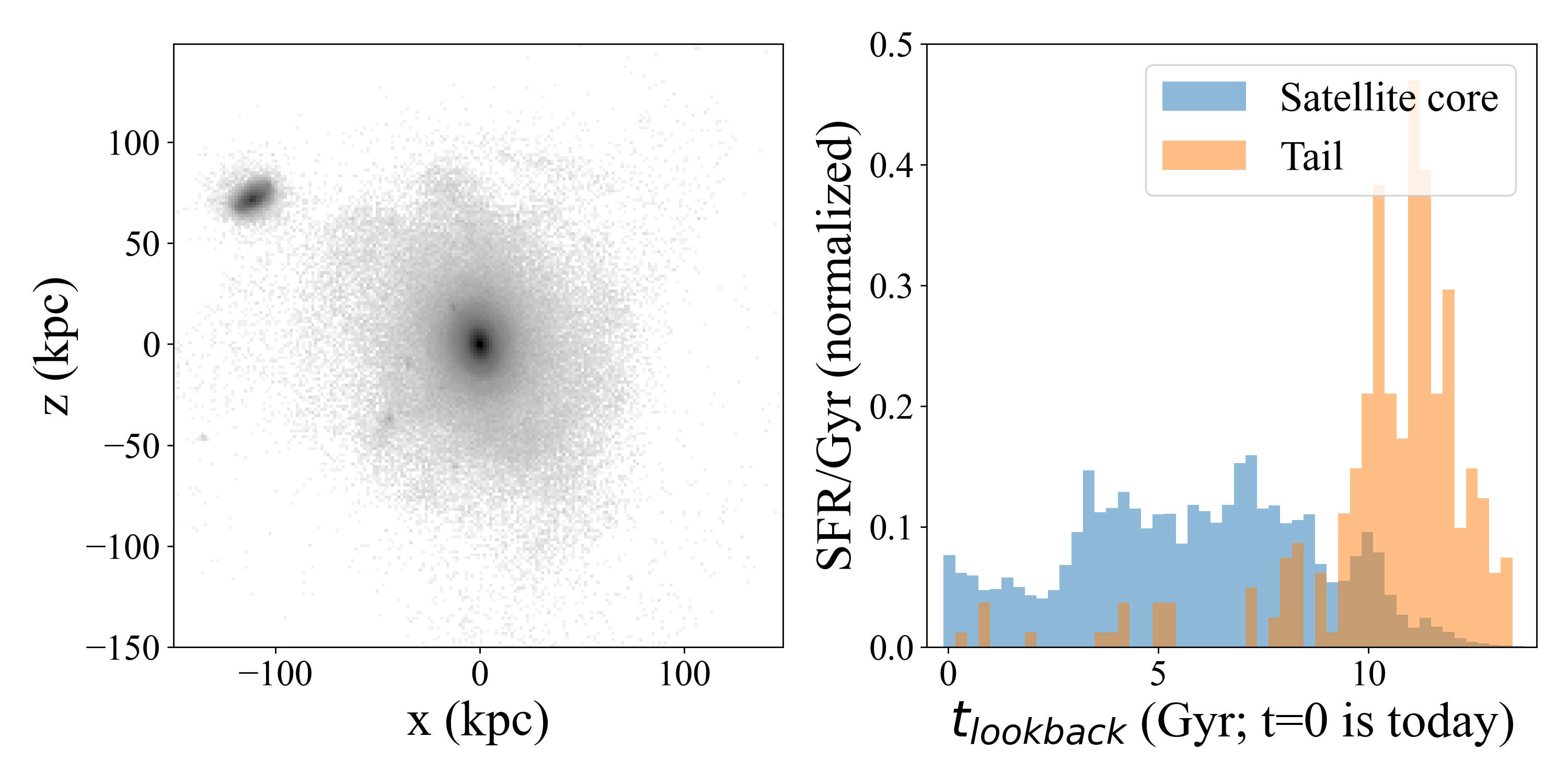}
    \caption{Two TNG50 analogs of NGC 3077 that have a similar mass, are orbiting around an M81-like host, and exhibit ongoing star formation in their cores. The upper panel shows a t$_{90}$ in the tidal tails of roughly 2 Gyr, while the lower exhibits a much older, 8 Gyr old population. The SFH of the NGC 3077 analogs' centers and analogs' tails are shown in the histograms on the right.}
    \label{fig:simsNGC}
\end{figure}

The outskirts of NGC 3077, in its morphology, large amount of unbound material, and continuous metallicity gradient from its inner parts to its outer parts, has long been argued to be tidal debris from its interaction with M81 \citep{Okamoto2015,Smercina2020,Okamoto2023}.
We confirm the metallicity measurements from \citet{Okamoto2023} of NGC 3077's tidal debris, and provide first constraints on the star formation history of the tidal arms, $t_{90} \sim 5.7$\,Gyr in the inner tidal stream, and $t_{90} \sim 3.6$\,Gyr in the outer tidal stream. No $t_{90}$ estimate exists for NGC 3077's main body, but recent star formation is seen in the central parts of NGC 3077 \citep{Harris2004}, suggesting that it is likely that it has a relatively recent $t_{90}$ (recall a constant SF history gives $t_{90}\sim1.3$\,Gyr). Yet, NGC 3077's pericenter passage with M81 is likely to have been $\sim250$\,Myr ago \citep[e.g.,][]{Yun1999}. One might have expected on that basis that NGC 3077's tidal tail could show a low $t_{90}$, of order 1.3--1.5\,Gyr, given that star formation could be shut off when the material was tidally stripped, but have been ongoing beforehand. Yet, NGC 3077's tails are older. Is it possible that this is because of age gradients in NGC 3077 prior to pericenter passage, or does this suggest that these tails may predate NGC 3077's most recent pericenter passage?

We explore this issue in a preliminary way using selected systems from the TNG-50 galaxy formation simulation. We choose a few example systems where a LMC-mass ($\log_{10} M_*/M_{\odot} \sim 9.5$) companion had a recent pericenter ($\sim 350$\,Myr ago in snapshot 97) with a roughly MW-mass ($\log_{10} M_*/M_{\odot}$ between 10.5 and 11.3) central galaxy. We illustrate the range in outcomes with two systems (Figure \ref{fig:simsNGC}) --- subhalo IDs 371127/371129 (central/satellite respectively) and 392276/392279. We calculate the $t_{90}$ values of the tidal tails from the LMC-mass companion. In the case of subhalo 371129, the tidal tail's $t_{90} \sim 1.8$\,Gyr shows a clear shut-off in star formation, while its core experiences enhanced tidally-induced star formation --- in line with our expectation that tails created by a recent pericenter passage should be young. In contrast, the tails of subhalo 392279 are much older, $t_{90} \sim 8.2$\,Gyr, while again the core of the satellite continues to form stars until the present day --- these tidal tails were formed in the recent pericenter passage 350\,Myr ago, but are comprised primarily of old stars, as the outer parts of this satellite (before its pericenter passage) were old (and much older than the core of the satellite). Neither of these timescales align precisely with those observed in NGC 3077 (3.6--5.7\,Gyr), but instead emphasize that the t$_{90}$ of tidal tails is not straightforwardly related to \emph{when} a merger event or tidal disruption happened (as suggested by \citealt{Harmsen2023}), especially in the case where satellites have strong population gradients prior to their tidal disruption. 

Recent work on Local Volume tidal streams has reached similar conclusions regarding population gradients present pre-disruption. The metallicity variations in the Sagittarius streams has been argued to imply strong population gradients in its progenitor galaxy \citep[e.g.][]{Keller2010,Hayes2020,Cunningham2024}. \cite{Keller2010} mention how the disruption of the Sagittarius progenitor galaxy is very recent, meaning the progenitor must have had inefficient mixing pre-disruption. Relatively consistent metallicity in the streams of NGC 3077 may imply efficient mixing in its past. Additionally, M33 is dominated by older stellar populations in its extremities \citep{Smercina2023} similarly to what we show here for NGC 3077 in that the tails are intermediate aged while the center is actively forming stars. Further investigations into the population gradients in NGC 3077's tidal tails in the near future will allow for more of its history to be further constrained.

\subsection{A Timeline of The M81 Group}
\label{sec:timeline}

\begin{figure*}[bth]
    \centering
    \includegraphics[width=\linewidth]{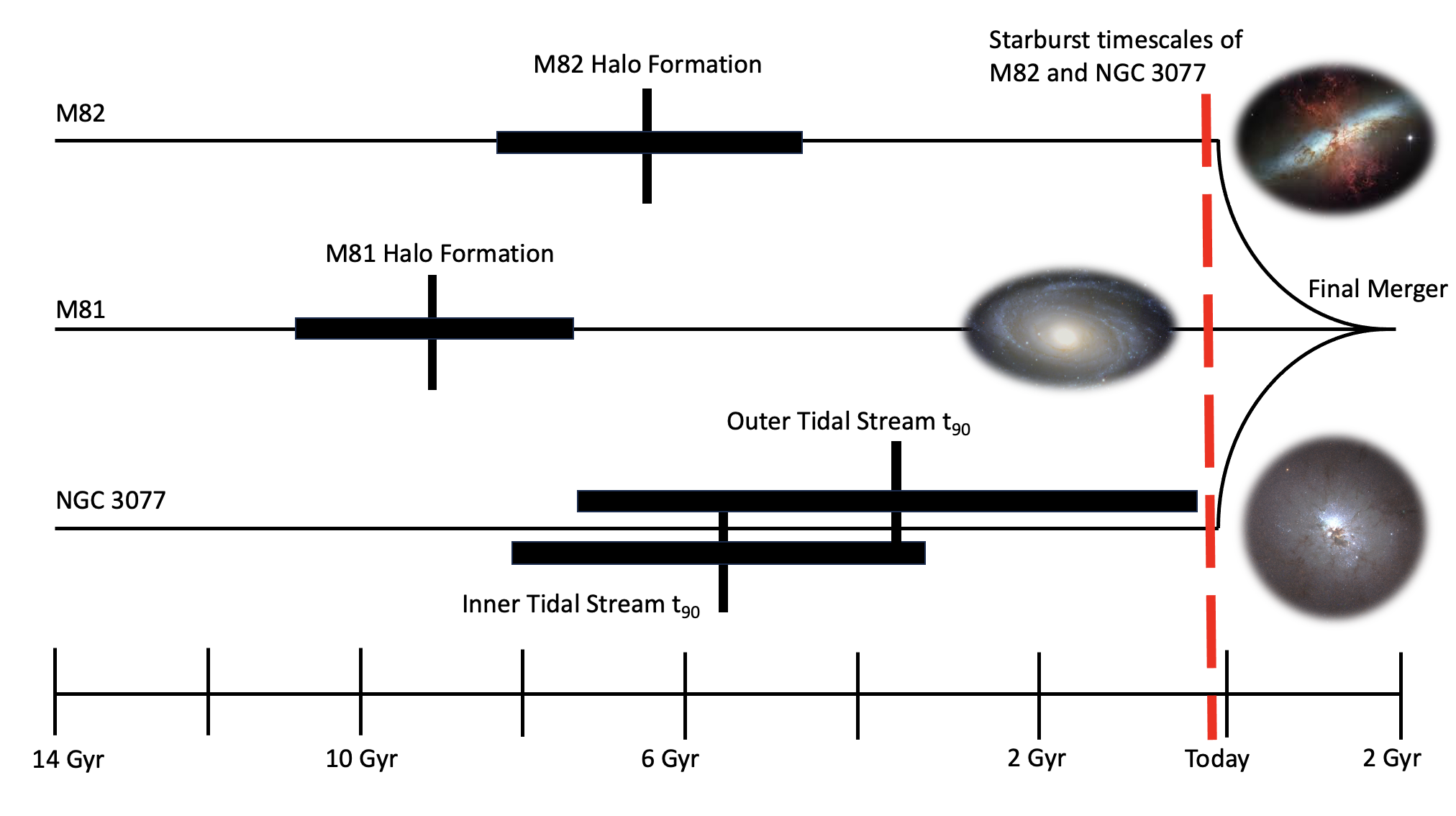}
    \caption{A timeline showing the star formation timescales in the outskirts of the central galaxies in the M81 group. The starburst timescales for M82 and NGC 3077 (red) are taken from various literature sources (NGC 3077: \cite{Yun1999}, M82: \cite{Smith2007, deMello2008}) which we adopt to be the time at which M82 and NGC 3077 experienced the strongest tidal forces, i.e., when disruption began. The final merger of these three galaxies will occur in 2 Gyr, as estimated by \cite{Smercina2020}.}
    \label{fig:timeline}
\end{figure*}

With new estimates of $t_{90}$ in hand for the envelopes of M82 and NGC 3077, we are in a position to reflect on how these new measurements help to enrich our understanding of the growth of the galaxies in the M81 group (and the group as a whole). 

We present a timeline of the M81 group in Figure \ref{fig:timeline}. M81's stellar halo formed early, $\sim9$\,Gyr ago, and has experienced little star formation until the present day \citep{Durrell2010}.  M82 appears to have a stellar halo, whose $t_{90}$ value of $\sim 6.6$\,Gyr suggests that it formed from a roughly SMC-massed accretion at around that time, substantially after M81's most important previous accretion. 

In contrast, the starburst timescales for M82 (200--300\,Myr, \citealt{deMello2008};  150--220\,Myr, \citealt{Smith2007}), cluster ages in NGC 3077 ($\sim$ 20\,Myr; \citealt{Harris2004}), and dynamical models of the interaction of M82 and NGC 3077 with M81 \citep[e.g.,][suggests an interaction timescale of $\sim$250\,Myr]{Yun1999} suggest that the periods of most intense tidal forces experienced by M82 and NGC 3077 were a few hundred Myr ago. These very recent timescales appear not to be reflected in the outskirts of either M82 --- whose outskirts appear to be a metal-poor stellar halo --- or NGC 3077 --- whose outskirts are tidal debris, but whose older ages appear to instead reflect a pre-existing population gradient that is now reflected in more recently formed tidal tails.  

Looking forward, it is clear that when NGC 3077 and (especially) M82 merge with M81 in a few Gyr, they will contribute to a massive, metal-rich halo with a substantial young stellar population \citep{Smercina2020}, both because of young stars in the main bodies of NGC 3077 and M82 that will contribute to the halo, but also because of the M81 group's widespread star formation in the gas tidal tails \citep[e.g.,][]{Okamoto2015,deBlok2018}. In this context, it is valuable to reflect on the relatively high $t_{90}$ values of NGC 3077's outskirts (and M82's halo) --- while star formation will be expected to continue well into M82's and NGC 3077's merger into the M81 group, the final stellar halo may have significant population gradients, where the outer parts (comprised of the debris liberated from the outer, older parts of the satellites) will likely be substantially older than the inner parts of the halo. Such a pattern is seen in M31's halo, where the tidal stream and inner halo have more younger stars than its outer halo (e.g., \citealt{Brown2008}, \citealt{DSouzanBell2018a}), perhaps also signifying that the progenitor of M31's halo had relatively older stars in its outer parts.

\subsection{Limitations and Future Work}
\label{sec:limitations_future}

The ideal way to constrain the time since a merger event is with a full SFH fit over a large footprint. Past work using HST data \citep{Durrell2010, Rejkuba2022} have had success finding star formation timescales in stellar halos and tidal streams and connecting them to merger history, but the small area these fields cover (or that would be covered by JWST) limits their ability to characterize the global properties of these faint structures. Therefore wide field, but shallow, resolved stellar population data are needed to perform the analysis necessary to constrain merger history. These methods, like the one used in this work, rely on an uncontaminated stellar sample; limiting contamination, both from background compact galaxies and removing foreground Milky Way dwarf stars from the AGB sample, is crucial for better-constrained measurements to be made. \citet{Rejkuba2022} shows that near-IR J $\minus$ K vs. K CMD characterization allows for foreground dwarf vs.\ giant separation, and the Nancy Grace Roman Space Telescope's \citep{Roman} combination of tight and well-characterized PSF, relatively deep limits, wide-field capability, and near infrared photometry will provide both excellent morphological star--galaxy separation and color--magnitude separation between foreground dwarf stars and the giant stars in the M81 group. 

Additionally, the current constraints on the relationship between AGB/TRGB and $t_{90}$ show considerable scatter \citep{Harmsen2023}; future work may refine our understanding of this relationship and/or uncover other relationships between the relative numbers of different luminous stellar evolutionary stages and other metrics of SFH. In addition, there is a clear need for an improved understanding of the relationship between SFH signatures and inferences about tidal interaction/merger history: the clear disconnect between $t_{90}$ and the growth timescale of NGC 3077's tidal tails exemplifies this concern. 

Finally, the last numerical model of this group is roughly 25 years old, based almost entirely on constraints from the {\sc Hi} tidal debris morphology \citep{Yun1999}. There is an urgent need for updated numerical modeling that incorporates the constraints given by the analysis of the resolved stellar populations (e.g., this work, \citealt{Okamoto2015, Okamoto2019,Smercina2020,Okamoto2023}) for a more nuanced interpretation of these and future results. 

\section{Conclusions}
\label{sec:conclusions}
We report the t$_{90}$ and [M/H] of the extended structures of M82 and NGC 3077. Using data from the Subaru Telescope's HSC, we have removed background and foreground contaminants to isolate the diffuse structures surrounding these galaxies. A large catalog of stars passed our strict requirements, and completeness corrections were done on our selections based on existing GHOSTS survey data. Spatial selections of the stellar halo of M82 and the eastern tidal stream of NGC 3077 were made, avoiding contamination from M81 and nearby tidal bridges.

Following \cite{Harmsen2023}, the ratio of AGB stars to TRGB stars was measured and converted to t$_{90}$ to determine when star formation stopped in our spatial selections. Our only deviation from the methods outlined by \cite{Harmsen2023} was the shortening of the AGB and TRGB CMD selection boxes to remove contamination from helium burning stars. Metallicity measurements for each of our selections were made following \cite{Ogami2024} for the sake of comparison with TNG50 analogs and literature sources. 

We confirm that M82 has a genuine stellar halo that formed before any interactions with the M81 group. It has global t$_{90}$ of $\sim$6.6 Gyr and a [M/H] of $\sim -$1.62 dex, signifying that it had very few major merger events, the most significant of which happening at the t$_{90}$ reported above. We hypothesize that the mass of the accreted satellite was roughly that of the SMC.

We also find that the uncontaminated inner tidal stream section (t$_{90}$ $\sim$5.8 Gyr) and outer tidal stream section (t$_{90}$ $\sim$3.6 Gyr) of NGC 3077 do not share an end to star formation, and the difference between them signifying a mildly negative (old to young) profile. We find that they share a metallicity (inner [M/H] $\sim -1.43$ dex, outer [M/H] $\sim -1.56$ dex) well within their errors ($\pm 0.20$ dex). The end to star formation we find does not align with its disruption timescale, and further complicates this galaxy's uncertain history. We suggest that NGC 3077 must have had a pre-existing age gradient in its outer disc before tidal disruption for any profile to be observable.

We show our results in the context of past work on M81's stellar halo and the starburst timescales of M82 and NGC 3077 to uncover the halo and tidal stream assembly timeline of this galaxy group. The halo of M82 is much younger than that of its massive host, and the difference in age between the stars in each section of NGC 3077's tidal tails is prominently displayed. None of the timescales we measure in this work align with the time at which M82 and NGC 3077 began to exhibit starburst behavior, which is the time we attribute to the beginning of their tidal disruption. The stellar populations we measure were therefore not formed in the fallout of this tidal disruption event; they predate the strong gravitational interactions these galaxies are currently experiencing and give insight into how the galaxies in this group came to be in their current states, as well has how this group will evolve in the future.

This work was partly supported by HST grant GO-16191 provided by NASA through a grant from the Space Telescope Science Institute, which is operated by the Association of Universities for Research in Astronomy, Inc., under NASA contract NAS5-26555. We acknowledge support from the National Science Foundation through grant NSF-AST 2007065 and by the WFIRST Infrared Nearby Galaxies Survey (WINGS) collaboration through NASA grant NNG16PJ28C through subcontract from the University of Washington. AM gratefully acknowledges support by FONDECYT Regular grant 1212046 and by the ANID BASAL project FB210003, as well as funding from the Max Planck Society through a “PartnerGroup” grant. This research has made use of NASA's Astrophysics Data System Bibliographic Services. 

Based on observations utilizing Pan-STARRS1 Survey. The Pan-STARRS1 Surveys (PS1) and the PS1 public science archive have been made possible through contributions by the Institute for Astronomy, the University of Hawaii, the Pan-STARRS Project Office, the Max-Planck Society and its participating institutes, the Max Planck Institute for Astronomy, Heidelberg and the Max Planck Institute for Extraterrestrial Physics, Garching, The Johns Hopkins University, Durham University, the University of Edinburgh, the Queen's University Belfast, the Harvard-Smithsonian Center for Astrophysics, the Las Cumbres Observatory Global Telescope Network Incorporated, the National Central University of Taiwan, the Space Telescope Science Institute, the National Aeronautics and Space Administration under Grant No. NNX08AR22G issued through the Planetary Science Division of the NASA Science Mission Directorate, the National Science Foundation Grant No. AST-1238877, the University of Maryland, Eotvos Lorand University (ELTE), the Los Alamos National Laboratory, and the Gordon and Betty Moore Foundation.

Based on observations obtained at the Subaru Observatory, which is operated by the National Astronomical Observatory of Japan, via the Gemini/Subaru Time Exchange Program. We thank the Subaru support staff --- particularly Akito Tajitsu, Tsuyoshi Terai, Dan Birchall, and Fumiaki Nakata --- for invaluable help preparing and carrying out the observing run. 

The authors wish to recognize and acknowledge the very significant cultural role and reverence that the summit of Maunakea has always had within the indigenous Hawaiian community. We are most fortunate to have the opportunity to conduct observations from this mountain.

\software{\texttt{HSC Pipeline} \citep{Bosch2018}, \texttt{Matplotlib} \citep{Hunter2007}, \texttt{NumPy} \citep{vanderWalt2011,Harris2020}, \texttt{Scipy} \citep{scipy}, \texttt{Astropy} \citep{astropy}}

\pagebreak

\bibliographystyle{aasjournal}
\bibliography{refs}

\appendix
\section{Removal of Potential Contaminants from CMD Selections}
\label{appendix_heb}
Here we show the original \citep{Harmsen2023} and shifted CMD selections used in this work on a CMD of the minor axis of M81. The isochrones are 100 Myr old helium burning stars with various metallicities. The shift on the blue edge of our selection (0.4 mag) was chosen to both avoid contamination from helium burning stars while still capturing the entirety of the Red Giant Branch. Using these CMD selections, we find the minor axis of M81 to have a $t_{90}$ of 8.6 $\pm$ 3.2 Gyr, matching the findings of \cite{Durrell2010} (9 $\pm$ 2 Gyr). The original selections give a $t_{90}$ of 5.8 $\pm$ 2.5 Gyr, illustrating the importance of removing HeB contaminants. 

\begin{figure}[bth]
    \centering
\includegraphics[scale=.55,keepaspectratio]{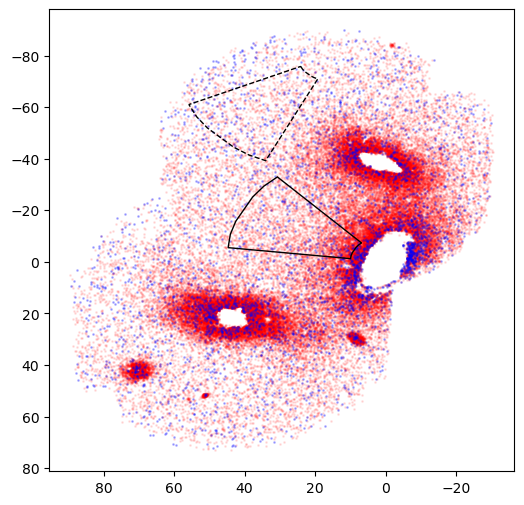}
    \caption{The minor axis and background selections of M81's minor axis shown in the same manner as those in Figure \ref{fig:spatial3}.}
    \label{fig:m81_appendix}
\end{figure}

\begin{figure}[bth]
    \centering
\includegraphics[scale=.45,keepaspectratio]{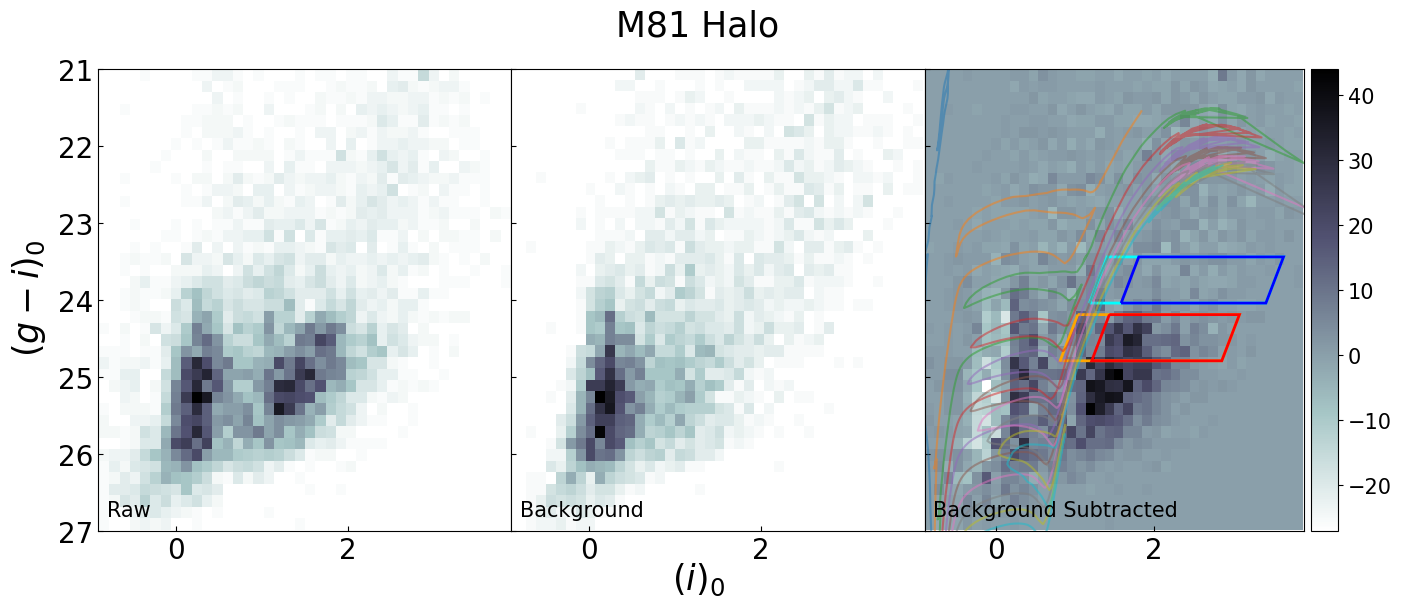}
    \caption{The raw, background, and background subtracted CMDs of the selections shown in Figure \ref{fig:m81_appendix} in the same manner as those in Figure \ref{fig:CMD_all}. The shifted selections are shown in red and blue, and the originals are in cyan and orange. The isochrones are 100 Myr Helium burning populations with various metallicities. }
    \label{fig:heb_shift}
\end{figure}

\end{document}